# Light matter in the core of the Earth: its identity, quantity and temperature using tricritical phenomena

**A Aitta**

Institute of Theoretical Geophysics, Department of Applied Mathematics and Theoretical Physics, University of Cambridge, Wilberforce Road, Cambridge CB3 0WA, UK

E-mail: A.Aitta@damtp.cam.ac.uk

**Abstract.** Light elements in the iron-rich core of the Earth are important indicators for the evolution of our planet. Their amount and distribution, and the temperature in the core, are essential for understanding how the core and the mantle interact and for modelling the geodynamo which generates the planetary magnetic field. However, there is a longstanding controversy surrounding the identity and quantity of the light elements. Here, the theory of tricritical phenomena is employed as a precise theoretical framework to study solidification at the high pressures and temperatures where both experimental and numerical methods are complicated to implement and have large uncertainties in their results. Combining the theory with the most reliable iron melting data and the Preliminary Reference Earth Model (PREM) seismic data, one obtains the solidification temperature at the inner core boundary (ICB) for both pure iron and for the alloy of iron and light elements in the actual core melt. One also finds a value of about 2.5 mole% for the amount of light matter. In addition, the density of both solid and liquid pure iron at its melting temperature is found. This allows one to obtain the density of the light matter and thus to identify it to be $MgSiO_3$.

## 1. Introduction

Although the Earth's core is impossible to reach directly, seismic wave scattering can provide information on the material properties in the deep layers of the Earth. By comparing these to the experimental high pressure studies of various materials, Birch in 1952 was the first to be able to conclude the composition of the Earth's core: "The inner core is most simply interpreted as crystalline iron, the outer part as liquid iron, perhaps alloyed with a small fraction of lighter elements."[1] Several light elements such as H, C, N, O, Mg, Si, P, S, K and some compounds of these have since been considered being present in the core, but the identity and quantity of the light matter has remained unresolved [1-8]. The seismic PREM data [9], known to an accuracy of about 1 %, demonstrates that the density difference between solid and melt at the Earth's ICB is greater than that expected from contraction of matter in solidification, indicating that the core melt has some light elements alloyed with iron, but the densities under high pressure for iron and light elements are much less well established. Direct experimental study of material densities at high pressures and temperatures is difficult and the estimates are based on either shock compression data, whose temperatures are not well-known, or extrapolations using thermal expansion measurements at much lower temperatures (see short summaries in [10, 11]). Theoretical simulations give approximate densities at these extreme pressures, but the results seem to be changing with improving techniques [12, 13]. The uncertainty of these results is increased by the unknown crystal structure which needs to be predetermined for these calculations, and the uncertainty of the iron melting temperature, which, as found by the same technique, in one case [14] is quoted to be about ± 600 K. Reliable and accurate estimates of the densities of iron and light elements at core pressures and temperatures are essential for success in identifying the core's light matter.

Here a complementary, precise theoretical approach is employed. The theory of tricritical phenomena, extending previous work [15] by further considerations, is used to give an improved estimate of the melting temperature of pure iron over the whole range of core pressures. From this the temperature of iron-rich melt as occurs at the ICB is found. One can then calculate the mole fraction of light matter in the melt, with no input from the iron density. Next, again using the theory of tricritical phenomena, the densities of pure iron, for both solid and melt at the melting temperature, with their uncertainties, are found, at core pressures from the constraint given by PREM on density at the ICB. Finally, one finds the density and the identity of the light matter.

## 2. The theory of tricritical phenomena for solidification

The density of matter under pressure increases until the distance between atoms becomes less than the atomic diameter; then atoms lose their individuality and the matter changes phase from liquid or solid to plasma where nuclei and electrons are free [16]. This transition from condensed matter to warm dense plasma has been suggested [15] to have a tricritical point where liquid, plasma and solid phases meet. In [15] the melting temperature of material whose melting curve as a function of pressure has a horizontal tangent, was investigated using Landau's theory of tricritical phenomena [16-17]. The theory of tricritical phenomena is applicable to a great variety of physical phenomena, including phase transitions between liquid and crystals [17], and it is known to be very accurate in problems in three spatial dimensions [18-19]. It is also shown to be quantitatively applicable along the phase transition line in a wide neighbourhood around a tricritical point [20]. In the case of a melting curve, the branching out of the vaporization curve and its end point are likely to have their own neighbourhoods outside the validity regime of the tricritical point and its critical line neighbourhood. However, this leaves a large pressure range from room pressure along the melting curve to beyond the tricritical pressure where the theory should be valid, excluding possibly the limited neighbourhoods at a few points where a different curvature can dominate locally in the presence of some triple points related to the transitions to different crystal structures below the melting curve.

One uses an order parameter $x$ to describe first order phase transitions which change to be second order at a tricritical point. $x = 0$ for the more ordered solid phase which occurs at lower temperature, and in the less ordered phases, the liquid or plasma, $x \neq 0$. The Gibbs free energy density is proportional to the Landau potential, which needs to be a sixth order polynomial in $x$:

$$\Phi(P,T;x) = \frac{1}{6} x^6 + \frac{1}{4} g(P) x^4 + \frac{1}{2} \varepsilon(T) x^2 + \Phi_0(P,T).$$ (1)

No higher order terms in $x$ appear in $\Phi$ since they can be eliminated using coordinate transformations as in bifurcation theory [21]. This method also scales out any dependence on physical parameters of the coefficient of the highest order term. When the melting curve has a horizontal tangent at high pressures, the coefficients $g$ and $\varepsilon$ can be taken to depend linearly on pressure $P$ and temperature $T$, respectively, and both are equal to 0 at the tricritical point ($P_{tc}$, $T_{tc}$). They are normalized to be

$$g(P) = \frac{P - P_{tc}}{P_{tc}} \quad \text{and} \quad \varepsilon(T) = \frac{3}{16} \frac{T_{tc} - T}{T_{tc} - T_0} \;, \tag{2}$$

where $T_0$ is the melting temperature at $P = 0$. The thermodynamically stable state occurs at the value(s) of $x$ that globally minimizes $\Phi$. For different $(P, T)$ values, $\Phi(x)$ can have one, two or three local minima. The temperature $T_m$ of the solid-liquid melting transition is determined by the condition that there are three minima of $\Phi$, all equally deep. This occurs when

$$\varepsilon = 3g^2/16 \text{ and } g < 0, \tag{3}$$

giving

$$T_m = T_{tc} - (T_{tc} - T_0)(P_{tc} - P)^2 / P_{tc}^{\;2}, \text{ for } P < P_{tc}. \tag{4}$$

Below $T_m$, the solid phase with $x = 0$ is the stable state, but the liquid phase (which corresponds to a local minimum with $x \neq 0$) can exist as an unstable state down to the temperature

$$T_U = T_{tc} - \frac{4}{3}(T_{tc} - T_0)(P_{tc} - P)^2 / P_{tc}^{\;2}, \text{ for } P < P_{tc} \tag{5}$$

which is where $\Phi$ changes from having three minima to having only a single minimum at $x = 0$, and there

$$\varepsilon = g^2/4 \text{ with } g < 0. \tag{6}$$

Stable liquid at temperatures lower than $T_m$ can be achieved using impurities: for instance in the familiar eutectic binary phase diagram the melting temperature of the mixture at any concentration is lower than the melting temperature for the closer pure end member. Impurities may stabilize the liquid all the way down to $T_U$. However, no negative temperature for $T_U$ is possible, as liquids will always solidify at or above absolute zero. Thus there is an absolute lower limit for $T_U$: at $P = 0$, $T_U = 0$. This gives

$$T_{tc} = 4T_0 \tag{7}$$

which simplifies (4) to

$$T_m = T_0 \, [4 - 3(P_{tc} - P)^2 / P_{tc}^{\;2}], \text{ for } P < P_{tc}, \tag{8}$$

and (5) to

$$T_U = 4T_0 [1 - (P_{tc} - P)^2 / P_{tc}^{\;2}], \text{ for } P < P_{tc}. \tag{9}$$

### *2.1. On the applicability to other materials*

This theory may be applied to other materials than iron which also have a horizontal tangent on their melting curve at high pressure. Unfortunately, there are not usually enough high pressure data to indicate clearly the curvature of the melting curve and its tangent at high pressures, but Cu, Mg, U and possibly Al seem to be promising candidates; however, all are in need of some more high pressure data to provide verification. For Al the curvature of its experimental melting curve at the moment depends crucially on a single shock wave data point whose value and uncertainty is a result of an interpretation of a figure in the original report [22] showing much wider range of uncertainty if one includes the choice for two Grüneisen parameter values than what is expressed in [23] or [24]. If the temperature for this data point were interpreted to be about 3750 K corresponding to a Grüneisen parameter value of 2.0 (as employed in [25] and [26], instead of 2.2 as in [23] and [24]), then it would appear that the melting curve has a horizontal tangent. Then equation (8) above, with the static data [23-24], gives a tricritical point at (97 GPa, 3734 K). Both sides of this tricritical point can be accessed experimentally for evidence of different behaviour on each side. An unexpected behaviour, interpreted as solid-liquid mixed phase, has already been reported [26] at pressures above 125 GPa.

### *2.2. Critical temperatures for pure iron*

Since $T_0 =1811$ K for pure iron, it follows from (7) that $T_{tc} = 7244$ K. The value for $P_{tc}$ can be estimated using the most reliable high-pressure data for $T_m$ (the same data [27-33] as selected earlier [15]). This gives $P_{tc} = (608 \pm 6)$ GPa, about 200 GPa less than the estimate found previously where both $T_{tc}$ and $P_{tc}$ were simultaneously determined without utilizing (7). Figure 1 shows the best fit (solid line) with its uncertainty (dashed lines) and the data with their error bars. Also, $T_U$ is shown in figure 1 (the lower solid line), with its uncertainty (dashed lines). From (8) and (9), one finds that at $P = 329$ GPa, the ICB pressure, $T_m = (6100 \pm 30)$ K and $T_U = (5710 \pm 40)$ K, whose uncertainties follow from the uncertainty in $P_{tc}$. This pure iron melting temperature $T_m$ is between the cluster of estimates around 6000 K obtained by various methods [34] and $(6350 \pm 600)$ K found in ab initio calculations [14]. $T_U$ determines the absolute lower limit down to which any impurities in the iron-rich core melt can lower the melting temperature. Its value is very close to 5700 K, the value inferred for the temperature at the ICB from ab initio calculations on the elasticity properties of the inner core [35], and is consistent with the range 5400 K – 5700 K reported in [14]. The temperature difference $T_m - T_U = (383 \pm 9)$ K at ICB is less than, for instance, the estimate 600 K to 700 K in [14] but more than the 300 K used in the rather recent energy budget calculations of the core [36].

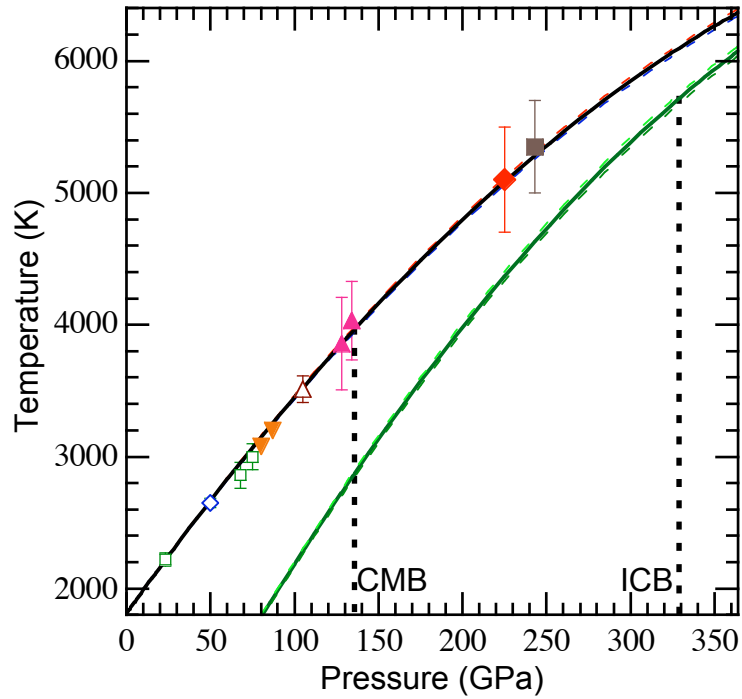

Figure 1. Iron melting temperature $T_m$ as a function of pressure $P$ as calculated from equation (8) (upper solid line) with its maximum uncertainty (dashed lines) and the consistent experimental data whose selection details are discussed in [15]. Open symbols are for the most accurate static measurements (squares [27]; diamond [28]; triangle [29]). Solid symbols show all the supporting shock wave results (downward triangles [30]; triangles [31]; diamond [32], square [33]. Error bars are from the original sources, but [30] gave none. The limit for liquid temperature, $T_U$, as a function of pressure $P$ as calculated from equation (9) (lower solid line) with its maximum uncertainty (dashed lines). The core boundary pressures are shown by vertical dotted lines.

### 2.3. Critical concentration at ICB

For a small (less than about 5 atomic %) amount of impurities, an iron-dominated alloy can be approximated by an ideal solution. Then the iron concentration $c_{Fe}$ can be related to the alloy's chemical potential and that of pure iron by $\mu_{alloy} = \mu_{Fe} + RT \ln c_{Fe}$ (see, for instance, [37]). On the other hand, chemical potential is Gibbs free energy per mole and thus proportional to the Landau potential (1): $\mu = \frac{128}{3} L_0 \Phi$, where the proportionality coefficient determined in previous work [15] is simplified by using (7) and the latent heat $L_0$ at zero pressure. When the liquid temperature equals $T_U$, $\Phi_{alloy} = \Phi_0$ (corresponding to having three equally deep minima in the potential) and $\Phi_{Fe} = \Phi_0 - g(P_{ICB})^3 / 48$ which follows from (6) and (1) with $x^2 = -g(P_{ICB})/2$. The concentration for the impurities is therefore $c_i = 1 - c_{Fe} = 1 - \exp\left[-\frac{8L_0}{9RT_U}\left(\frac{P_{tc} - P_{ICB}}{P_{tc}}\right)^3\right]$ giving $c_i = (2.47 \pm 0.07)$ mole%, where the uncertainty corresponds to the maximum uncertainty in $P_{tc}$ and $T_U$. Thus, by this method, one can find the concentration of light elements at the ICB without using the iron density, which is not known very accurately. The small fraction of Ni expected also in the core is here assumed not to change the iron behaviour due to the remarkable similarities of these atoms.

### 2.4. Critical densities at ICB

In order to identify the light matter, the iron melt density needs to be approximated. The theory of the tricritical point gives an equation similar to (4) for the density of the liquid at its melting temperature, $\rho_m = \rho_{tc} - (\rho_{tc} - \rho_0)(P_{tc} - P)^2 / P_{tc}^2$, for $P < P_{tc}$, where $\rho_0$ is the zero pressure liquid density and $\rho_{tc}$ is the liquid density at $P_{tc}$, both at their melting temperature. For pure iron, the liquid state exists as an unstable state relative to the solid for densities down to

$\rho_U = \rho_{tc} - \frac{4}{3}(\rho_{tc} - \rho_0)(P_{tc} - P)^2 / P_{tc}^2$, for $P < P_{tc}$. Below $\rho_U$ only solid state is possible. However, light elements dissolved into the iron-rich core melt can generate densities less than that of pure iron. Here $\rho_U$ is identified to be the density of the light-element enriched iron melt at ICB and its value $\rho_U = \rho_m^{PREM}(P_{ICB})$ is known from PREM (cf. $T_U$ was identified to be the temperature there). This gives, with $P_{tc}$ found above, $\rho_{tc} = (14.20 \pm 0.14)$ g/cm$^3$, where the uncertainty corresponds to the range obtained using $\pm 1$ % uncertainty in PREM combined with the uncertainty in $P_{tc}$. This density is consistent within the uncertainties of the similar densities at high pressure for iron's principal Hugoniot in [38], [39] and the older data tabulated in [40] even though the temperatures they correspond to are not known and need not be the melting temperature. According to [41] the very high-pressure measurements (beyond 620 GPa in [42], that is, beyond $P_{tc}$ found here) extend into the regime of increasing thermal ionization. Figure 2 presents $\rho_m$ as a thin solid line and $\rho_U$ as a dot-dashed line, together with the PREM density data as a thicker solid line. In particular, at ICB, $\rho_m = (12.67 \pm 0.07)$ g/cm$^3$, about 4.1 % higher than $\rho_m^{PREM}$ and 0.7 % lower than the PREM density of the solid $\rho_s^{PREM}$.

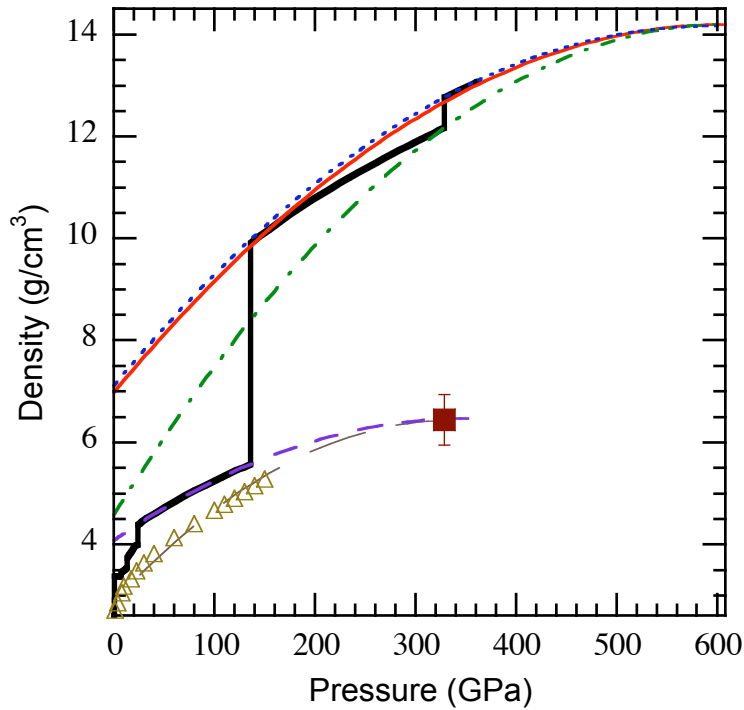

Figure 2. Densities as a function of pressure up to $P_{tc}$. Density of solid iron $\rho_s$ (dotted line) and liquid iron $\rho_m$ (thin solid line) at their melting temperature with the limit for liquid density $\rho_U$ (dashed-dotted line) and PREM [9] density (thick solid line). Square shows the density found here for the light matter MgSiO₃ in the melt at the inner core boundary. To demonstrate its reasonability it is connected by quadratic fits to both PREM lower mantle densities (dashed line) and to the new data [53] (triangles) for liquid MgSiO₃ density at its melting temperature (long dashed line).

One also finds that at the CMB, $\rho_U$ is close to 1 % below $\rho_m^{PREM}$, indicating that the core melt hardly has light impurities there or some Ni compensates them in the melt. This is about 8 % less than obtained from ab initio calculations (see, for instance, [13]) which at that pressure also give about 10 % (or 6 % with free energy correction) greater temperatures for the melting [12, 14]. But the value of $\rho_U$ is in full agreement with the similarity of compressional wave velocity measurements against density in pure Fe and seismic velocity profile against density from PREM at the pressures of the upper outer core [43]. This means that light matter brought to CMB by convection is removed from the melt by solidification accompanied by some iron. From (8) the melting temperature $T_m$ at the CMB pressure is $(3965 \pm 20)$ K, very similar to its seismic estimate $(3950 \pm 200)$ K [44]. This is further evidence that at the CMB the core melt is solidifying, as has been suggested in [45-47]. Even though there is presently no experimental evidence concerning the solidification temperature of relevant matter from iron-rich melt at the CMB pressure, the temperature must be close to the pure iron melting temperature owing to the very small amount of impurities.

The volume contraction (from [15] with (7) and using $L_0$) in solidification, $\Delta V = 6L_0 (P_{tc} - P)^2 / P_{tc}^3$, is $(0.0288 \pm 0.0004)$ cm³/mol for iron at ICB pressure, which is less than an estimate of 0.055 cm³/mol used for $\varepsilon$–iron in [48]. One can find the density $\rho_s$ of pure solid iron from the density of the melt $\rho_m$ using $\rho_s = A/(A/\rho_m - \Delta V)$, where the atomic weight $A$ is taken to be the standard value for iron on the Earth. $\rho_s$ is shown in figure 2 as a dotted line. In particular, at ICB, $\rho_s = (12.758 \pm 0.006)$ g/cm³ whose uncertainty only takes account of the uncertainties in $\rho_m$ and $\Delta V$. The upper end of the uncertainty range includes $\rho_s^{PREM}$ indicating that the inner core is very nearly pure iron. These results give a smaller than traditionally expected $(10 \pm 5)$ % density deficit relative to pure iron but rather close to the prediction of 5.4 % in [48].

### 3. Light matter content, density and identification

The average density of the impurities $\rho_i$ at ICB can be estimated using the equation $\frac{1}{\rho_m^{PREM}} = \frac{y}{\rho_i} + \frac{1-y}{\rho_m}$, where $y$ is the weight fraction of the impurities, valid strictly for homogeneous solutions but probably also for the fluid in the core due to the very small amount of impurities present and the very long time the core melt has been mixed by convection. In figure 3, $y$ and $\rho_i$ are presented as a function of molar mass for $c_i = 2.47$ mole% as was obtained earlier. No single light element has a high enough $\rho_i$ [40] in agreement with previous research [4, 7, 14] which suggests that the light matter needs to have at least two elements. The previously suggested two likely elements O and Si [14, 49-51] if combined as $SiO_2$ also generates too low $\rho_i$ [40]. Thus it becomes important to expand the search to the common three-element compounds, particularly the most abundant mantle compounds forsterite, $Mg_2SiO_4$, and magnesium silicate, $MgSiO_3$. If the compound were forsterite, one obtains $\rho_i = 7.5$ g/cm³ which seems to be too high [52-53], but if it were $MgSiO_3$, then $\rho_i = (6.5 \pm 0.5)$ g/cm³, and this result is very reasonable [53]. Even though the density of any light element or compound at ICB conditions is not known accurately, using this latter value of $\rho_i$ one can find a smooth second order continuation of the new density estimates in [53] for liquid $MgSiO_3$ at its melting temperature. This $\rho_i$ can also be quadratically aligned with the PREM lower mantle densities of solid material which is believed to be dominated by $MgSiO_3$. $\rho_i$ for $MgSiO_3$ and these two fits are shown in figure 2. For $MgSiO_3$, $y = (4.36 \pm 0.13)$ wt% implying that at ICB the melt has about 1.06 wt% of Mg, 1.22 wt% Si and 2.09 wt% O. The amount of the light matter found by this method is less than the recent results in [14] and [43], however, the iron density used in those calculations is over 2 % higher than that found here. This implies their need to include some light elements in the inner core, too.

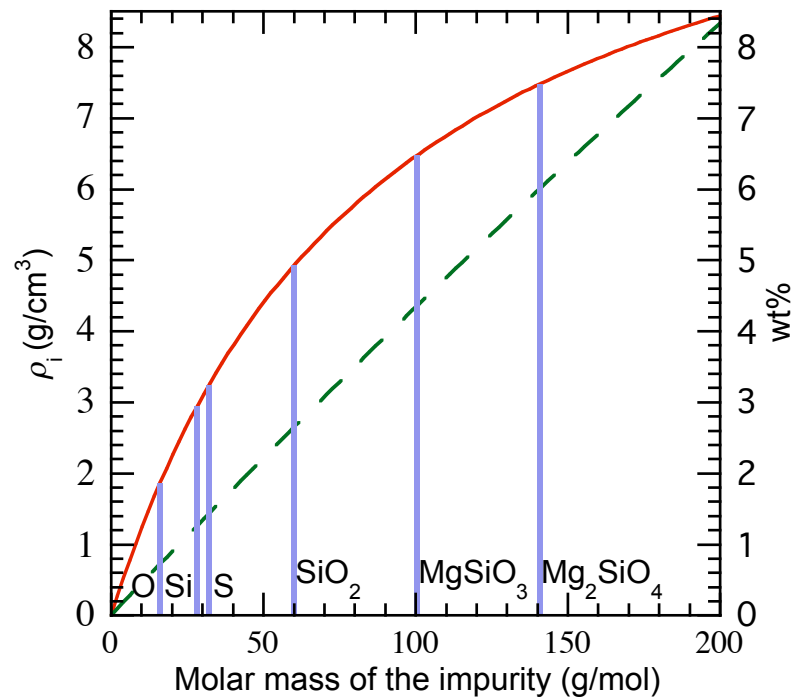

Figure 3. Density and weight fraction of the impurities as a function of molar mass for the obtained impurities' mole fraction 0.0247. To guide the eye, the mole fractions for O, Si, S, $SiO_2$ and the main mantle components $MgSiO_3$ and $Mg_2SiO_4$ are shown by vertical lines.

The light matter found here suggests that in the partially or fully molten early Earth, a small quantity of the molten main mantle material accompanied the iron when it descended as a melt to the core. The compound $MgSiO_3$ has in the past [49, 54] been considered as the core's light compound, and so have each of its elements. In particular, Si and O, alone or together, have been often suggested as light elements in the core. Magnesium has been suggested more rarely: it was suggested by Alder in 1966 [54] but later discounted on account of the weak solubility of MgO in iron melt even at high pressures [4]. However, recently Mg has been found adequately soluble in iron at high pressures and thus suggested as an important light element in the core [55]. The result found here calls for further investigations of the liquid density of Fe, Ni and $MgSiO_3$, both pure and as alloys, at core conditions. It will also be worth investigating the crystal structure and the Fe-concentration of the solid, and especially the solidification temperature, as a small amount of molten $MgSiO_3$ in Fe, or in Fe-Ni, is cooled down at the CMB pressure.

## 4. Conclusions

The theory of tricritical phenomena provides a concise framework to calculate precisely the quantities essential to understand the Earth's core. Accurate quantitative predictions are obtained for the iron melting temperature and melting density at high pressures, and for the temperature and the amount and identity of the light matter in the real Earth at the inner core boundary. The density of the solid inner core is found to be similar to pure iron. The density of light matter in the melt corresponds to $MgSiO_3$. Thus the core consists of the four most abundant elements of the Earth in addition to the commonly expected Ni not considered here separately. It is likely that the small amount of $MgSiO_3$ originates from the molten early mantle. This result agrees with the understanding that O is the principal light element in the core, and the expectation that the core has some Si. Additionally this work identifies the third main light element to be Mg. This challenges the present geochemical calculations where no O or Mg is assumed in the core. Furthermore, this study indicates that the light matter solidifies out at CMB with some iron, in agreement with previous suggestions.

**Acknowledgement**

I thank Paul Asimow for providing me with reference [53] in a tabulated form.